\documentclass[amsmath, amssymb,10pt,aps,prb,twocolumn,notitlepage,showpacs,superscriptaddress]{revtex4-2}
\usepackage{amsmath}
\usepackage{graphicx}
\usepackage{calc}
\usepackage{braket}
\usepackage{ulem}   
\usepackage{slashed}
\usepackage{wasysym}
\usepackage{siunitx}
\usepackage{dsfont}
\usepackage{amsthm,amsmath,amsfonts,amssymb,verbatim,color}
\usepackage{graphicx}
\usepackage{subfigure}
\usepackage{bm}
\usepackage{epsfig,slashed}
\usepackage[T1]{fontenc}
\usepackage[colorlinks=true,citecolor=blue,linkcolor=blue,urlcolor=blue]{hyperref}
\usepackage[version=4]{mhchem}
\begin{document}
\title{Observation of terahertz second harmonic generation from Dirac surface states in the topological insulator Bi$_2$Se$_3$}

\author{Jonathan Stensberg}
\affiliation{Department of Physics and Astronomy, University of Pennsylvania, Philadelphia, Pennsylvania 19104, U.S.A}
\author{Xingyue Han}
\affiliation{Department of Physics and Astronomy, University of Pennsylvania, Philadelphia, Pennsylvania 19104, U.S.A}
\author{Zhuoliang Ni}
\affiliation{Department of Physics and Astronomy, University of Pennsylvania, Philadelphia, Pennsylvania 19104, U.S.A}
\author{Xiong Yao}
\affiliation{Department of Physics and Astronomy, Rutgers, The State University of New Jersey, Piscataway, New Jersey 08854, U.S.A.}
\author{Xiaoyu Yuan}
\affiliation{Department of Physics and Astronomy, Rutgers, The State University of New Jersey, Piscataway, New Jersey 08854, U.S.A.}
\author{Debarghya Mallick}
\affiliation{Department of Physics and Astronomy, Rutgers, The State University of New Jersey, Piscataway, New Jersey 08854, U.S.A.}
\author{Akshat Gandhi}
\affiliation{Department of Physics and Astronomy, Rutgers, The State University of New Jersey, Piscataway, New Jersey 08854, U.S.A.}
\author{Seongshik Oh}
\affiliation{Department of Physics and Astronomy, Rutgers, The State University of New Jersey, Piscataway, New Jersey 08854, U.S.A.}
\author{Liang Wu}
\email{liangwu@sas.upenn.edu}
\affiliation{Department of Physics and Astronomy, University of Pennsylvania, Philadelphia, Pennsylvania 19104, U.S.A}

\date{\today}

\begin{abstract}
We report the observation of second harmonic generation with high conversion efficiency $\sim 0.005\%$ in the terahertz regime from thin films of the topological insulator Bi$_2$Se$_3$ that exhibit the linear photogalvanic effect, measured via time-domain terahertz nonlinear spectroscopy and terahertz emission, respectively. As neither phenomena is observable from topologically trivial In-doped  Bi$_2$Se$_3$, and since no enhancement is observed when subject to band bending, the efficient thickness-independent nonliear responses are attributable to the Dirac fermions of topological surface states of Bi$_2$Se$_3$. 
This observation of intrinsic terahertz second harmonic generation in an equilibrium system unlocks the full suite of both even and odd harmonic orders in the terahertz regime and opens new pathways to probing quantum geometry via intraband second-order nonlinear processes. We hope our work will motivate the theoretical development of a full treatment of second harmonic generation for probing the quantum geometry in various inversion-breaking topological and twisted materials. 
 \end{abstract}

\maketitle

\textbf{Introduction} 

Harmonic generation (HG) has been an invaluable nonlinear optical technique since its first demonstration \cite{Franken1961} and continues to power recent advances ranging from the imagining of microscopic magnetic domains \cite{Ni2021NatureNano,Ni2021PRL} to the development of tabletop sources of extreme ultraviolet and x-ray light for attosecond science \cite{Luu2015,Li2020}. 
Nevertheless, employing HG to study phenomena below $\sim$ 100 meV has been severely impeded by the historical terahertz (THz) gap \cite{Tonouchi2007}, traditionally $\sim$ 0.1-30 THz (1 THz $\approx$ 4.1 meV), where technical challenges have impeded the development of intense light sources. Recent progress in intense THz generation \cite{Hoffmann2011,Hafez2016}, however, has enabled the first applications of HG to the THz regime. Since its first demonstration \cite{Matsunaga2014}, THz third harmonic generation (THG) has rapidly become a standard tool for characterizing the Higgs mode \cite{Matsunaga2014,Matsunaga2013,Tsuji2015,Pekker2015,Shimano2020} and other nonlinear optical processes \cite{Cea2016,Tsuji2016,Matsunaga2017,Moor2017,Yang2018,Yang2019,Nakamura2019,Seibold2021,Schwarz2021,Luo2022,Yuan2024,Feng2023,Kaj2023,Chu2023} in a variety of superconductors \cite{Kovalev2021PRB,Reinhoffer2022,Isoyama2021,Vaswani2021, Rajasekaran2018,Chu2020,Katsumi2023,Schwarz2020}. Yet more recently, odd-order THz-HG has been reported in doped Si \cite{Meng2020,Meng2022,Meng2023} and materials hosting Dirac fermions, namely graphene \cite{Hafez2018,Kovalev2021SciAdv,Deinert2021,Tielrooij2022}, Cd$_3$As$_2$ \cite{Cheng2020,Kovalev2020,Germanskiy2022}, and the bismuth chalcogenide family of topological insulators \cite{Giorgianni2016,Kovalev2021,Tielrooij2022}. The latest studies have explored effectively tuning the nonlinear process of THz-HG via gating \cite{Kovalev2021SciAdv} and metasurfacing \cite{Deinert2021,Tielrooij2022}.

Despite these advances, THz-HG remains highly constrained, limited almost exclusively to odd-order low harmonics. 
The most striking limitation is that even-order THz harmonics (which are only permitted under inversion symmetry breaking) have only been observed 
in superconductors with a net propagating supercurrent \cite{Vaswani2020,Nakamura2020} that drives the system out of equilibrium to break inversion symmetry, or in carefully engineered meta-material devices \cite{Lee2020} that artificially break inversion symmetry. Intrinsic even-order THz-HG has never been demonstrated in an equilibrium material. 
This is a severe impediment to studying quantum materials, as second harmonic generation (SHG) has physically different origins than the commonly-studied odd-order harmonics, including conventional asymmetric scattering \cite{Belinicher1980}, skew scattering of chiral Bloch fermions \cite{Isobe2020,He2022}, and quantum geometry \cite{Morimoto2016,Wu2017,Ma2021}. In the THz regime, interband transitions are precluded in most systems due to the low energy scale, and scattering contributions are diminished due to the frequency exceeding the scattering rate \cite{Konig2017,Matsyshyn2019}, making THz-SHG an ideal probe of the intraband quantum geometry 
(though a full derivation of intraband SHG from quantum geometry still needs to be developed) 
As prototypical topological insulators with a centrosymmetric bulk and inversion symmetry-breaking topological surfaces, the bismuth chalcogenides \cite{Valdes-Aguilar2012,Wu2013,Wu2015,Wu2016} provide an ideal platform to probe the intraband quantum geometry of the topological surface state via THz-SHG while intrinsically avoiding the properties of the bulk band. 
However, previous nonlinear THz studies of the bismuth chalcogenides \cite{Giorgianni2016,Kovalev2021,Tielrooij2022} failed to observe THz-SHG, 
despite the ubiquity of surface SHG--and even-order HG in general--
in the nonlinear optical response outside of the THz regime \cite{Hsieh2011a,Hsieh2011b,McIver2012PRB,Mizrahi1988,Bai2021,Schmid2021,Heide2022}. Realizing THz-SHG originating from Dirac surfaces states in the bismuth chalcogenides is therefore a crucial step toward opening a powerful new nonlinear pathway to measuring intraband quantum geometry via THz-HG techniques. 

Here, we report the observation of THz-SHG from Bi$_2$Se$_3$ samples exhibiting the linear photogalvanic effect (LPGE). With LPGE determined by THz emission \cite{Kampfrath2013} and THz-SHG measured via intense time-domain THz spectroscopy (TDTS) \cite{NussOrenstein1998}, thin films of Bi$_2$Se$_3$ that display LPGE are found to produce THz-SHG that is highly efficient and independent of the sample thickness. As these phenomena are not observed in topologically trivial In-doped Bi$_2$Se$_3$, and since there is no enhancement observed when samples are subject to band bending, the efficient response is attributable to the Dirac fermions of the topological surface states. We further show that as both LPGE and SHG result from second-order nonlinear processes, both effects originate from unequal populations of twinned domains of the three-fold symmetric surface of Bi$_2$Se$_3$, which motivates the future development of techniques to preferentially control the orientation of crystal growth on millimeter scales, particularly for materials that break various symmetries. 
This observation of intrinsic SHG in the THz regime in an equilibrium system 
enables the future THz-HG investigation of nonlinear intraband quantum processes via the full suite of harmonic orders, both even and odd. \\

\textbf{Results and Discussion}

Thin film samples of Bi$_2$Se$_3$ are grown via molecular beam epitaxy on c-axis Al$_2$O$_3$ substrates (10 mm x 10 mm x 0.5 mm), following standard growth procedures, utilizing the two-step growth process \cite{Bansal2011,Bansal2012}. 
As each van der Waals unit of Bi$_2$Se$_3$ forms a quintuple layer (QL), samples of 16 QL, 32 QL, 64 QL, and 100 QL are grown to form a thickness series (1 QL $\approx$ 1 nm). Band bending can occur at the surface of Bi$_2$Se$_3$ \cite{Hsieh2009,Brahlek2011,Valdez-AguilarJAP2012}, 
which introduces an internal electric field $E_{int}$ near the surface and results in an effective $\chi^{(2)}$ for second-order processes of $\chi_{eff}^{(2)} = \chi_{ijk}^{(2)} + \chi_{ijkl}^{(3)}E_{int}$ \cite{Hsieh2011a,McIver2012PRB,Zhu2015}. 
Therefore, the samples are capped \textit{in situ} with 50 nm of Se, which greatly mitigates band bending, isolates the intrinsic second-order response, and protects against damage \cite{Salehi2015}. 
Finally, as sufficient In-doping suppresses the topological surface states of Bi$_2$Se$_3$ while being subject to the same band bending effects \cite{Wu2013,Giorgianni2016}, a Se-capped 100 QL sample of (Bi$_{0.9}$In$_{0.1}$)$_2$Se$_3$ is grown to directly discriminate between topological and trivial origins.

\begin{figure}
    \centering
    \includegraphics[width=\columnwidth]{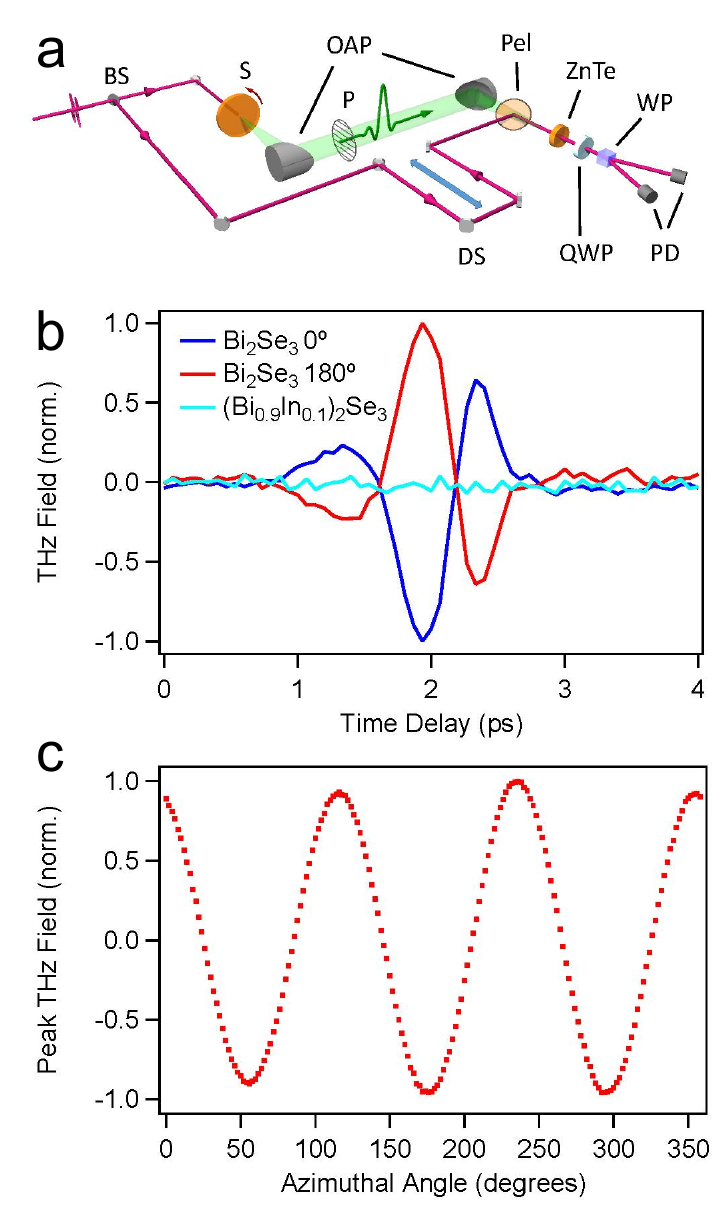}
    \caption{\textbf{a.} Schematic of the THz emission spectrometer. The NIR and THz beam paths are depicted in magenta and green, respectively, and the THz beam path is contained in a dry air-purged box. Both sample (S) and polarizer (P) are mounted in rotating stages to enable characterization of the azimuthal angle dependence of the THz emission. Labeled optical elements include beam splitter (BS), pelical (Pel), ZnTe crystal (ZnTe), quarter wave plate (QWP), Wollaston prism (WP), photodiodes (PD), and delay stage (DS). \textbf{b.} Normalized electric field profile of the emitted THz pulse from Se-capped 100 QL Bi$_2$Se$_3$ and (Bi$_{0.9}$In$_{0.1}$)$_2$Se$_3$ obtained by electro-optic sampling mapped in the time domain. \textbf{c.} The peak normalized electric field as the Bi$_2$Se$_3$ azimuthal angle $\phi$ is rotated.}
    \label{fig:Fig1}
\end{figure}

\begin{figure*}
    \centering
    \includegraphics[width=\textwidth]{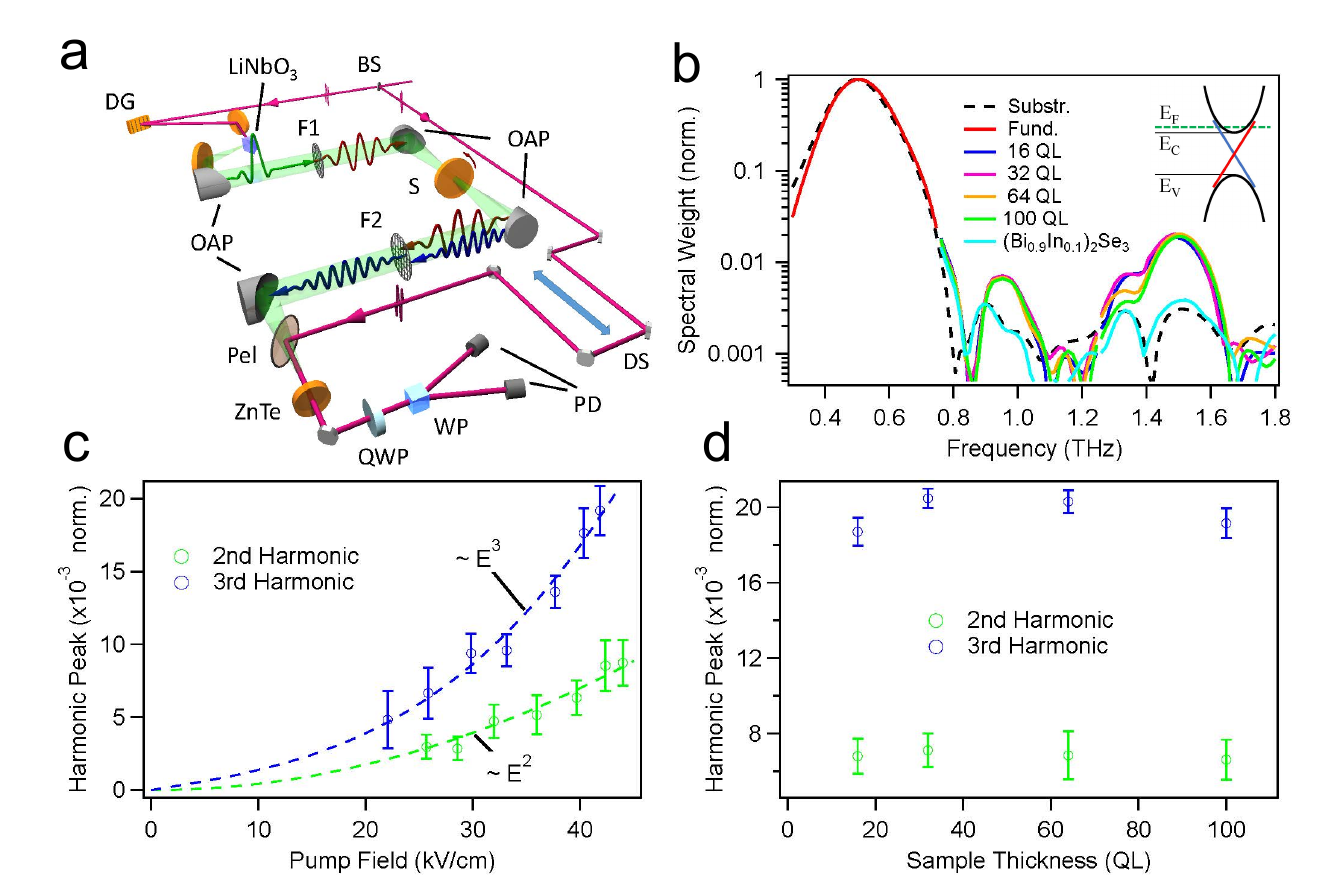}
    \caption{\textbf{a.} Schematic of the intense TDTS system. The NIR and THz beam paths are depicted in magenta and green, respectively, and the THz beam path is contained in a dry air-purged box. Labeled optical elements include sample (S), LiNbO$_3$ crystal (LiNbO$_3$), THz filters (F1 and F2), diffraction grating (DG), beam splitter (BS), pelical (Pel), ZnTe crystal (ZnTe), quarter wave plate (QWP), Wollaston prism (WP), photodiodes (PD), and delay stage (DS). \textbf{b.} Harmonic generation spectra for Se-capped Bi$_2$Se$_3$ and (Bi$_{0.9}$In$_{0.1}$)$_2$Se$_3$ samples under a 0.5 THz fundamental pump with respect to a reference substrate. The change between spectra taken with 1.0 THz-specific and 1.5 THz-specific filters are indicated by breaks in the spectra, and the Bi$_2$Se$_3$ band structure is sketched with the Fermi level (inset). \textbf{c.} Bi$_2$Se$_3$ peak spectral weight at the 2nd and 3rd harmonic as a function of the peak 0.5 THz pump field $E_{pump}$, with fits to $E_{pump}^2$ and $E_{pump}^3$ respectively. \textbf{d.} Peak spectral weight at the 2nd and 3rd harmonics as function of sample thickness.}
    \label{fig:Fig2}
\end{figure*}

The samples of Bi$_2$Se$_3$ are evaluated for their room temperature LPGE response by measuring the THz emission \cite{Kampfrath2013} of the samples under normal incidence, near infrared (NIR) pumping.  
When a single domain of Bi$_2$Se$_3$ is pumped above the bulk band gap of $\sim$ 0.3 eV, 
 photocurrent current generation moves free carriers across the domain \cite{McIver2012NatNano,Zhu2015,Braun2016}. These carriers are generated from the so-called shift vector from quantum geometry, and also scatter asymmetrically off of the wedge-shaped potential of the trigonal domains \cite{Belinicher1980,Olbrich2014},  yielding a low-frequency LPGE current with directionality defined by the domain orientation and which couples out to free space as a THz pulse. This emitted THz pulse is generated and detected by a THz emission spectrometer depicted schematically in Fig \ref{fig:Fig1}.a and described in previous works \cite{Ni2020,Ni2021}. In brief, the sample is pumped over a spot size of order 1 mm by linearly polarized, broadband 1530 nm (0.81 eV), 50 fs pulses with a repetition rate of 1 kHz. A quasi-single cycle THz pulse is emitted from the sample in transmission geometry; collected, collimated, and focused onto a ZnTe crystal by a pair of off-axis parabolic mirrors in $4f$ geometry; and measured via electro-optic sampling \cite{Wu1996}. By varying the optical path length of the NIR probe pulse via the delay stage, the electric field profile of the emitted THz pulse $E_{THz}$ is mapped out in the time domain. 

THz emission data is depicted in Fig \ref{fig:Fig1}.b,c for typical 100 QL Bi$_2$Se$_3$ and (Bi$_{0.9}$In$_{0.1}$)$_2$Se$_3$ samples. As shown in \ref{fig:Fig1}.b, the (Bi$_{0.9}$In$_{0.1}$)$_2$Se$_3$ emits no signal whereas a pronounced quasi-single cycle THz pulse is emitted from the Bi$_2$Se$_3$, 
the polarity of which changes sign throughout the duration of the pulse when the sample is rotated azimuthally by 180 degrees. By tracing out the peak value of $E_{THz}$ as the sample is rotated, as shown in Fig \ref{fig:Fig1}.c, the azimuthal angle dependence clearly follows $E_{THz}^{max} = E_0\sin{(3\phi+\phi_0)}$, where $E_0$ is the peak electric field strength, $\phi$ is the azimuthal angle, and $\phi_0$ is an arbitrary angle difference between the crystalline axes and the lab frame for a given sample. 

\begin{figure*}
    \centering
    \includegraphics[width=\textwidth]{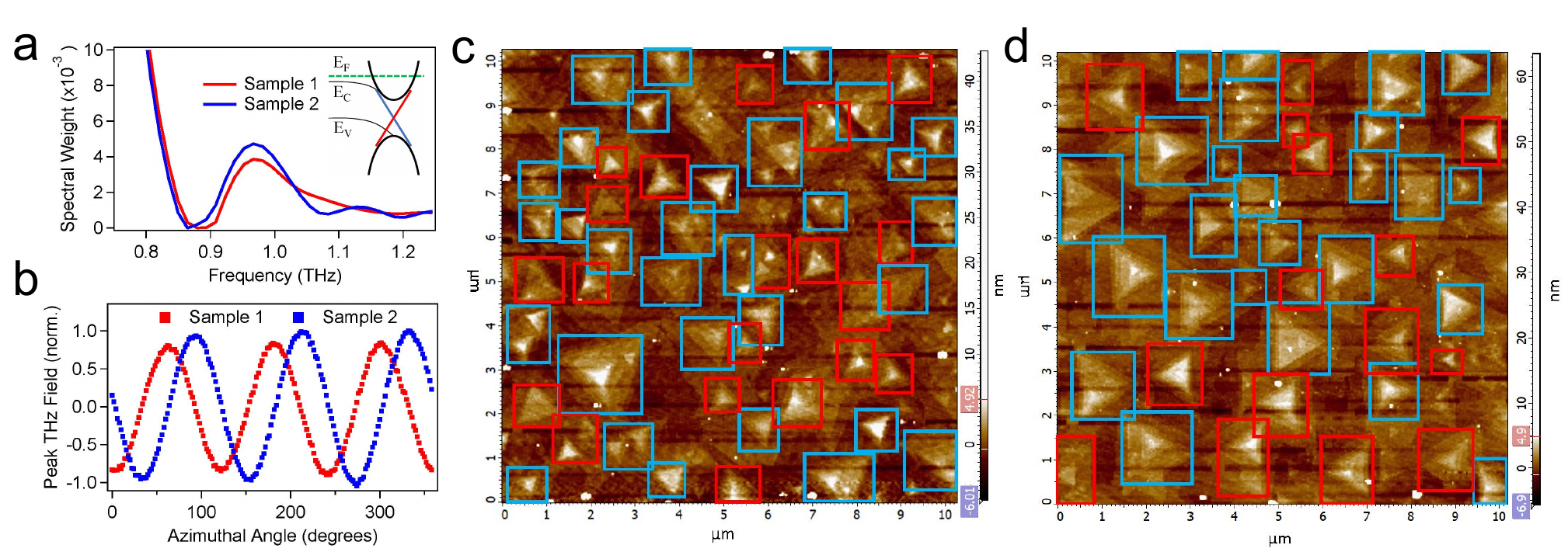}
    \caption{\textbf{a,b.} Comparison of THz-SHG and LPGE, respectively, for two bare 100 QL Bi$_2$Se$_3$ samples. The band bending effect on the uncapped samples is sketched in (\textbf{a}, inset), and the azimuthal angle in (\textbf{b}) is offset for clarity. \textbf{c,d.} Atomic force microscopy images of representative areas of bare 100 QL Bi$_2$Se$_3$ for Sample 1 and Sample 2, respectively, where oppositely-oriented domains on the surface are highlighted with blue and red boxes.}
    \label{fig:Fig3}
\end{figure*}

This $\sin{(3\phi + \phi_0})$ dependence of the emitted $E_{THz}$ is precisely the azimuthal angle dependence expected for THz emission from a single domain of Bi$_2$Se$_3$ due to LPGE under normal incidence \cite{Hsieh2011a,Hsieh2011b,McIver2012PRB,Mizrahi1988}. LPGE is only permitted in systems that break inversion symmetry \cite{Belinicher1980}. As bulk Bi$_2$Se$_3$ is centrosymmetric, only the three-fold symmetric surface of Bi$_2$Se$_3$ breaks inversion and two-fold rotation symmetries 
and contributes to the LPGE, 
yielding a $\sin{(3\phi + \phi_0)}$ dependence of the LPGE current for a single domain under normal incidence (See Supplemental Material \cite{SM} for derivation). 
Since the spot size of the NIR pump (order 1 mm) vastly exceeds the domain size of Bi$_2$Se$_3$ (order 1 $\mu$m; see Fig \ref{fig:Fig3}.c,d), the THz emission method measures the net LPGE produced by a large ensemble of Bi$_2$Se$_3$ domains. Since twinned domains in the sample produce oppositely-signed LPGE responses, as demonstrated in Fig \ref{fig:Fig1}.b, the azimuthal dependence of the single domain LPGE can be generalized to the case of many domains as $(f_+ - f_-)\sin{(3\phi + \phi_0)}$, where $f_+ + f_- = 1$ indicate the fraction of oppositely-oriented domains, respectively \cite{SM}. 
The observation of a clear three-fold LPGE signal from the  Bi$_2$Se$_3$ therefore indicates the presence of a dominant domain orientation over millimeter length scales, such that $f_+ \neq f_-$.

As SHG is limited by the same symmetry considerations as LPGE and expected to be generated from the surface of Bi$_2$Se$_3$ \cite{Hsieh2011a,Hsieh2011b,McIver2012PRB,Mizrahi1988}, the THz-HG of the samples is measured via intense TDTS \cite{NussOrenstein1998} at room temperature as shown schematically in Fig \ref{fig:Fig2}.a. Intense broadband, quasi-single cycle THz pulses are generated from LiNbO$_3$ via the tilted pulse front method \cite{Hebling2002,Fulop2010,Kunitski2013} by pumping with linearly polarized, broadband 800 nm, 35 fs pulses with a repetition rate of 1 kHz. The generated intense THz pulses are collected, directed through the sample at a waist of order 1 mm, and focused onto a ZnTe crystal by a quartet of OAPs in $8f$ geometry. Prior to the sample, optical filters (F1) convert the broadband pulse into a narrow-band few cycle pulse centered at 0.5 THz (spectral width $\sim$~20$\%$). After transmitting through the sample, the resulting THz pulse is passed through optical filters (F2) to suppress the spectral weight of the 0.5 THz fundamental pulse and pass the frequency range around the harmonic to be observed: 1.0 THz for SHG or 1.5 THz for THG. The remaining THz that impinges upon the ZnTe crystal is measured by standard electro-optic sampling \cite{Wu1996}, allowing the electric field profile to be mapped out in the time domain by varying the delay stage of the probe pulse. Finally, taking the Fourier transform of the THz pulse in the time domain yields the spectral weight of the pulse as a function of frequency.

The HG spectra for the Bi$_2$Se$_3$ samples shown in Fig \ref{fig:Fig2}.b exhibit clear THz-SHG at 1.0 THz and THz-THG at 1.5 THz when pumping with the 0.5 THz fundamental; however, no HG above the background level is observed for (Bi$_{0.9}$In$_{0.1}$)$_2$Se$_3$. The common dip in the spectra at 1.4 THz is due to residual water absorption \cite{vanExter1989}. Such THz-HG has been attributed \cite{Hafez2018,Kovalev2021SciAdv,Deinert2021,Tielrooij2022,Cheng2020,Kovalev2020,Germanskiy2022,Giorgianni2016,Kovalev2021,Bai2021,Schmid2021,Heide2022} to the ultrafast dynamics of Dirac fermions as they are accelerated by the THz light field through the surface Dirac cone. This Dirac fermion origin of the efficient THz-HG in the bismuth chalcogenides is confirmed here by direct comparison with the topologically trivial (Bi$_{0.9}$In$_{0.1}$)$_2$Se$_3$ sample, where the THz-HG is dramatically reduced in the absence of Dirac fermions. Furthermore, three key features 
demonstrate strong agreement with the THz-THG observed in previous studies \cite{Giorgianni2016, Kovalev2021,Tielrooij2022} of bismuth chalcogenides:  First, the THG conversion efficiency is $\sim 0.04\%$ (accounting for the THG-specific filters), which closely matches the conversion efficiency in previous reports. Second, the yield of the THz-THG scales perturbatively as $E_{pump}^3$, as shown in Fig \ref{fig:Fig2}.c which likewise agrees with previous results and contrasts sharply with the saturation 
observed in graphene \cite{Hafez2018,Kovalev2021SciAdv,Deinert2021,Tielrooij2022} and Cd$_3$As$_2$ \cite{Cheng2020,Kovalev2020}. 
Third, the THz-THG yield is nearly thickness-independent, as shown in Fig \ref{fig:Fig2}.d, which is consistent with 
the dominant contribution to the THz-THG 
originating from the topological surface state. Together, these 
results reaffirm the conclusions of the previous studies, indicate that the intrinsic nonlinear properties of the Bi$_2$Se$_3$ samples measured here are consistent with those of the previous studies, and demonstrate the dominance of the surface Dirac fermions in the nonlinear THz response even when bulk contributions are not symmetry-forbidden.
 
Returning to Fig \ref{fig:Fig2}.b, a clear THz-SHG peak is observed at 1.0 THz for the Bi$_2$Se$_3$ samples, in addition to the THz-THG peak at 1.5 THz. As shown in Fig \ref{fig:Fig2}.c, the 1.0 THz peak scales according to the $E_{pump}^2$ expectation for a perturbative second-order response (See \cite{SM} for comparison to linear fit). Furthermore, the 1.0 THz peak is thickness independent, as 
shown in Fig \ref{fig:Fig2}.d, matching the expectation for Bi$_2$Se$_3$ where only the surface breaks inversion symmetry and two-fold rotation symmetry as required for a second-order process. Since bulk contributions are symmetry-forbidden \cite{Hsieh2011a,Hsieh2011b,McIver2012PRB,Mizrahi1988}, the band bending contribution is suppressed by the Se capping layer \cite{Salehi2015} (sketched in Fig \ref{fig:Fig2}.b inset) irrespective of the sample topology,  Dirac fermions are known to dominate the nonlinear THz response when present \cite{Giorgianni2016, Kovalev2021, Tielrooij2022,Hafez2018,Kovalev2021SciAdv,Deinert2021,Tielrooij2022,Cheng2020,Kovalev2020,Germanskiy2022,Bai2021,Schmid2021,Heide2022}, and no THz-SHG is observed from the topologically trivial Se-capped (Bi$_{0.9}$In$_{0.1}$)$_2$Se$_3$, this clear THz-SHG response from Bi$_2$Se$_3$ can be attributed to the topological surface state. While both the THz-SHG and THz-THG originate from the topological surface state, the yield of twinned domains cancels for THz-SHG according to $(f_+ - f_-)$ whereas the yield adds for THz-THG according to $(f_+ + f_-)$. Thus the lower THz-SHG yield compared to THz-THG in Fig \ref{fig:Fig2} accords with the theoretical expectation that twinned domains suppress the THz-SHG. This observation of THz-SHG reaching a high conversion efficiency of  $\sim 0.005\%$ (accounting for the SHG-specific filters) is consistent with HG studies outside of the THz regime \cite{Hsieh2011a,Hsieh2011b,McIver2012PRB,Mizrahi1988,Bai2021,Schmid2021,Heide2022}, but contrasts sharply with the previous THz studies \cite{Giorgianni2016, Kovalev2021,Tielrooij2022} of bismuth chalcogenides, which failed to report THz-SHG. 

Finally, we verify these conclusions by measuring capless Bi$_2$Se$_3$ samples. 
Fig \ref{fig:Fig3}.a,b respectively compare the THz-SHG and THz emission results for two 100 QL bare Bi$_2$Se$_3$ samples taken from the same batch to ensure similar growth quality, with both samples exposed to atmosphere for a similar duration of several hours to ensure stable and comparable $\chi_{eff}^{(2)}$ due to band bending \cite{Hsieh2011a,McIver2012PRB,Bai2021,Zhu2015,Braun2016} (sketched in Fig \ref{fig:Fig3}.a inset). 
The capless samples exhibit no enhancement of either LPGE or THz-SHG compared to the Se-capped samples, confirming that the band bending contribution is small and that the large second-order nonlinear response in Bi$_2$Se$_3$ is indeed due to the Dirac fermions. 

The lack of a cap permits the orientation of surface domains to be determined by atomic force microscopy (AFM). Various points on the samples were therefore measured by AFM to verify the presence of unequal populations of twinned domains. Fig \ref{fig:Fig3}.c,d respectively display representative AFM images with twinned domains clearly visible on the surface of both samples. 
As variations in domain size and quality average out over many domains, the ratio of many domains yields a rough but sufficiently accurate proxy for comparing the relative populations of oppositely-oriented domains. 
A careful counting of these domains shows that the ratio of oppositely-oriented domains is $\sim1.5:1$ in Sample 1 and $\sim1.8:1$ in Sample 2, demonstrating that both samples have unequal (not $1:1$) populations of twinned domains. These domain ratios correlate with the magnitude of the second-order responses for the samples, agreeing with the theoretical expectation and confirming the dependence of both LPGE and THz-SHG upon the presence of unequal populations of twinned domains. 
This dependence of the second-order processes upon unequal populations of twinned domains presents one potential reason why the previous studies \cite{Giorgianni2016, Kovalev2021,Tielrooij2022} of bismuth chalcogenides failed to report THz-SHG, and it highlights the importance of improving control over crystal growth 
for materials that break various symmetries.

To summarize, we have observed THz-SHG from the topological surface state of Bi$_2$Se$_3$ featuring a highly efficient conversion rate of $\sim 0.005\%$ and have proposed a plausible reason for the failure to observe THz-SHG in previous studies \cite{Giorgianni2016, Kovalev2021,Tielrooij2022} of similar bismuth chalcogenides, namely the suppression of second-order processes by twinned domains \cite{Giorgianni2016, Kovalev2021,Tielrooij2022}.
By demonstrating intrinsic SHG in the THz regime for an equilibrium system, these results open a new nonlinear pathway to experimentally probing intraband quantum geometry via THz-HG methods. This crucial experimental advance motivates the extensive theoretical elaboration of nonlinear intraband quantum processes, as has been accomplished for the interband \cite{Ma2021}, and enables further development of THz-HG techniques 
to discriminate between various nonlinear quantum effects. While various scattering \cite{Belinicher1980,Isobe2020,He2022} and quantum geometry \cite{Ma2019,Kang2019,He2021,Gao2023,Wang2023} contributions to intraband nonlinear Hall effect involving Dirac fermions have been proposed, a full treatment of intraband SHG is lacking \cite{Isobe2020} and remains for future theoretical work. As the experimental capacity to probe intraband SHG in the THz regime has now been demonstrated, developing such theoretical models is the primary extant barrier to the unexplored frontier of nonlinear intraband quantum processes. Altogether, these results vastly expand the range of future studies by unlocking even-order HG in the THz regime and open new experimental and theoretical pathways to the low-energy study of topological surface states and nonlinear intraband quantum processes in noncentrosymmetric topological semimetals and twisted material systems in order to study the quantum geometrical effects.\\

\textbf{Acknowledgement}

We thank F. de Juan, A. Grushin, T. Morimoto, and J. Lu for helpful discussions. J.S.  was initially sponsored by the Army Research Office under the YIP award to L.W. under Grant No. W911NF-19-1-0342 (completed in 2022), and was partially supported by the Gordon and Betty Moore Foundation’s EPiQS Initiative under the grant GBMF9212 to L.W. (completed in 2023), and is supported by the Dissertation Completion Fellowship  at  the  University  of  Pennsylvania. X.H. is  supported  by the NSF EPM program under grant no. DMR-2213891. Z.N. acknowledges support from  the Dissertation Completion Fellowship at the University of Pennsylvania. The work at Rutgers by X. Yao, X. Yuan, D. M., A. G. and S. O. were supported by the Army Research Office  under the grant No. W911NF-20-2-0166, NSF DMR2004125, and the center for Quantum Materials Synthesis (cQMS), funded by the Gordon and Betty Moore Foundation’s EPiQS initiative through grant GBMF10104. L.W. acknowledges the support by Air Force Office of Scientific Research under award no. FA955022-1-0410.

\bibliography{bi2Se3_bib.bib}

\end{document}